\documentclass[conference,10pt]{IEEEtran}
\usepackage{cite}
\usepackage{amsmath,amssymb}
\usepackage{graphicx}
\usepackage{array}
\usepackage{booktabs}
\usepackage{multirow}
\usepackage{xcolor}
\usepackage{balance}
\usepackage{rtospro}
\usepackage{url}
\usepackage[hidelinks]{hyperref}

\begin{document}

\title{DS2Agent: Holistic Device Agents for Accelerated Embedded Board-Level Design}


\author{%
  \IEEEauthorblockN{%
    Hongbin Yang\IEEEauthorrefmark{1},
    Chi Zhang\IEEEauthorrefmark{1},
    Yinhao Xiao\IEEEauthorrefmark{2},
    Runyu Pan\IEEEauthorrefmark{1},
    Yanni Yang\IEEEauthorrefmark{1}, and
    Xiuzhen Cheng\IEEEauthorrefmark{1}%
  }
  \IEEEauthorblockA{%
    \IEEEauthorrefmark{1}Shandong University\\
    \{hongbinyang, chi\_zhang\_2024\}@mail.sdu.edu.cn,
    \{rypan, yanniyang, xzcheng\}@sdu.edu.cn
  }
  \IEEEauthorblockA{%
    \IEEEauthorrefmark{2}GDUFE-UCM Joint Institute, Guangdong University of Finance \& Economic\\
    20191081@gdufe.edu.cn
  }
}

\maketitle
\begin{abstract}
Embedded board-level development demands extensive expertise, from component selection and device characteristics to interface configuration, driver writing, and system composition.
Recently, LLM-based agents have shown strong capabilities, raising the prospect of automating board-level design tasks.
However, existing agents either target chip-level design or treat datasheets as static text, offering little guidance on whether a specific component should---or even could---be used for a given board-level design.
Moreover, without touching real hardware, they hallucinate incorrect details.

To fill this gap, we present {\setlength{\fboxsep}{1pt}\href{https://anonymous.4open.science/r/DS2Agent-9FE8/}{\fcolorbox{blue}{white}{DS2Agent}}}, an open-source framework that constructs interactive agents from device datasheets.
Each agent comprises three models: a knowledge model supporting retrieval and question answering; a digital-twin model capturing the physical functionality of the device; and an interface model supporting driver development and system integration.
To suppress hallucinations, we also allow optional connections to real devices, so that the agent can calibrate its internalized knowledge against the device.
We evaluate DS2Agent on ten devices spanning seven functional categories.
Compared with the baseline, the DS2Agent-augmented agent improves the question-answering score from 87 to 97 out of 100 while using 44.8\% fewer tokens.
For driver generation, it reduces average token usage per device by 72.9\%.
For system integration, it reduces total token usage across three FreeRTOS applications by 29.7\%.
\end{abstract}


\section{Introduction}
\label{sec:intr}
Board-level embedded design requires engineers to select components, reconcile operating and interface constraints, develop drivers, and integrate them into a working system~\cite{lin2020polymorphic,lin2024boarddse}.
These decisions span system architecture and individual components, and often depend on experience with the device family, target platform, and application.
Even an experienced engineer may need to investigate an unfamiliar device before deciding whether it fits the design.

A manufacturer datasheet provides the device specifications needed for this investigation, including electrical characteristics, register maps, communication protocols, timing requirements, and packaging.
Engineers use these specifications to assess candidate devices before ordering parts or committing to a board design.
However, the information needed for a single decision can be spread across prose, tables, and timing diagrams, making it time-consuming to locate and interpret~\cite{zhu2026pcbqa}.
Engineers must then translate the relevant specifications into design constraints, firmware operations, and expectations about system behavior.
In practice, they need to answer several kinds of questions:

\begin{figure}[t!]
  \centering
  \includegraphics[width=\linewidth, trim=0 0 0 0, clip]{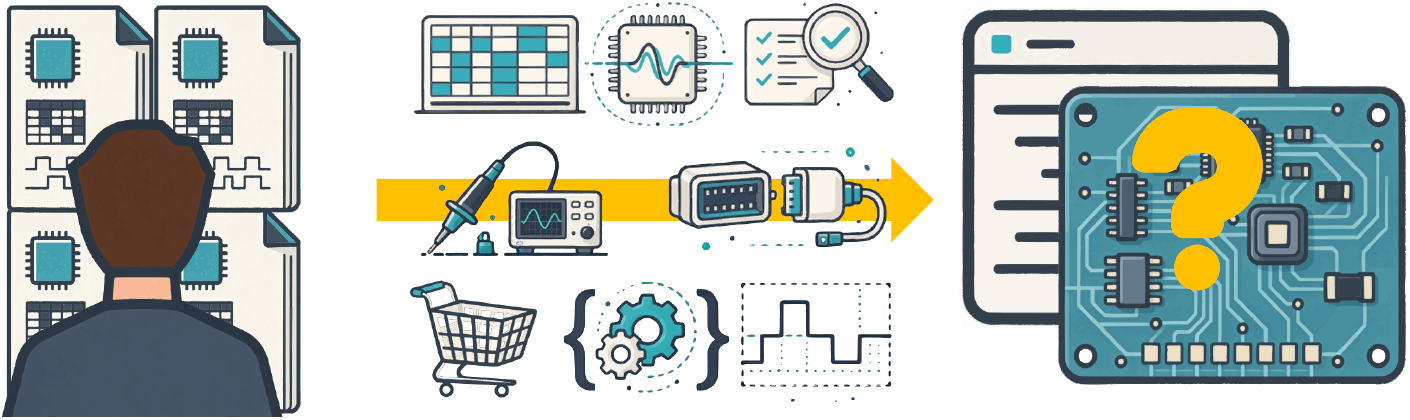}
  \par\smallskip
  {\footnotesize (a) Limitations of conventional and prototyping: tedious and laborious\par}
  \medskip
  \includegraphics[width=\linewidth, trim=0 0 0 0, clip]{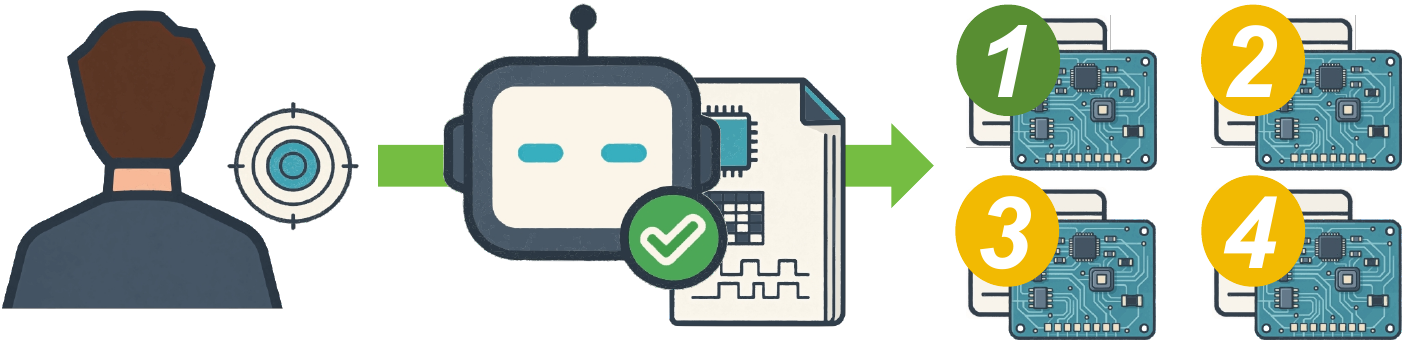}
  \par\smallskip
  {\footnotesize (b) Advantages of the agent-driven method: rapid and efficient exploration\par}
  \caption{Conventional and DS2Agent-assisted workflows for board-level embedded design.}
  \label{fig:effect}
\end{figure}

\definecolor{mygreen}{RGB}{0,168,98}
\definecolor{myred}{RGB}{198,40,40}
\begin{table}[t!]
  \scriptsize
  \centering
  \setlength{\tabcolsep}{3pt}
  \caption{Capability comparison with selected tools.}
  \label{tbl:compare}
  \begin{tabular}{|l|c|c|c|c|c|}
    \hline
    \makecell{Work} & \makecell{Board\\Level} & \makecell{Digital\\Twin} & \makecell{Hardware\\Calibration} & \makecell{MCP\\Interface} & \makecell{Design\\Exploration} \\
    \hline
    Arduino MCP~\cite{arduinoMCP} & \textcolor{mygreen}{\cmark} & \textcolor{myred}{\xmark} & \textcolor{myred}{\xmark} & \textcolor{mygreen}{\cmark} & \textcolor{mygreen}{\cmark} \\
    \hline
    Embedded MCP~\cite{embeddedMCP} & \textcolor{mygreen}{\cmark} & \textcolor{myred}{\xmark} & \textcolor{myred}{\xmark} & \textcolor{mygreen}{\cmark} & \textcolor{mygreen}{\cmark} \\
    \hline
    Labgrid~\cite{labgrid} & \textcolor{mygreen}{\cmark} & \textcolor{myred}{\xmark} & \textcolor{myred}{\xmark} & \textcolor{myred}{\xmark} & \textcolor{mygreen}{\cmark} \\
    \hline
    Renode~\cite{renode} & \textcolor{mygreen}{\cmark} & \textcolor{mygreen}{\cmark} & \textcolor{myred}{\xmark} & \textcolor{myred}{\xmark} & \textcolor{mygreen}{\cmark} \\
    \hline
    Eclipse Ditto~\cite{eclipseDitto} & \textcolor{myred}{\xmark} & \textcolor{mygreen}{\cmark} & \textcolor{myred}{\xmark} & \textcolor{myred}{\xmark} & \textcolor{myred}{\xmark} \\
    \hline
    OpenCalib~\cite{openCalib} & \textcolor{myred}{\xmark} & \textcolor{myred}{\xmark} & \textcolor{mygreen}{\cmark} & \textcolor{myred}{\xmark} & \textcolor{mygreen}{\cmark} \\
    \hline
    \textbf{DS2Agent} & \textcolor{mygreen}{\cmark} & \textcolor{mygreen}{\cmark} & \textcolor{mygreen}{\cmark} & \textcolor{mygreen}{\cmark} & \textcolor{mygreen}{\cmark} \\
    \hline
  \end{tabular}
\end{table}

\begin{itemize}
  \item \textbf{Functionality.} What does the device do, and which part of the design requirement does it satisfy?
  \item \textbf{Operating conditions.} What supply, temperature, and other environmental conditions does it require?
  \item \textbf{Device-specific behavior.} What restrictions, mode-dependent effects, or initialization requirements could affect the design?
  \item \textbf{Interface.} Which communication protocols and timing constraints must the host support?
  \item \textbf{Integration.} How should the firmware initialize the device, exchange data with it, and handle its responses?
  \item \textbf{Performance.} Under the intended configuration and operating conditions, will the device meet the system's performance requirements?
  \item \textbf{Availability.} Can the part be sourced in the required region and quantity at an acceptable price?
\end{itemize}

The datasheet informs the technical questions, while availability requires information from suppliers.
Engineers must weigh these factors together because a part that meets a functional requirement may still impose unacceptable integration costs or operating constraints.
With several candidates, the same analysis must be repeated for each device.
Experienced engineers can often anticipate integration problems early, but some details become apparent only when they implement the driver or test the assembled system.
For example, an overlooked timing constraint can invalidate a polling strategy and force changes to the firmware or the part selection.
Figure~\ref{fig:effect} contrasts this iterative workflow with the workflow supported by DS2Agent.

Existing tools address different stages of this process.
Board-level design tools use reusable circuit abstractions and enable engineers to explore component alternatives~\cite{lin2020polymorphic,lin2024boarddse}.
Large language models (LLMs) have also been applied to chip design and RTL generation~\cite{liu2023chipnemo,liu2025rtlcoder}.
At the board level, PCB-Bench evaluates LLMs on placement, routing, and design understanding, while PCBAgent addresses high-density PCB component placement~\cite{li2026pcbbench,chen2025pcbagent}.
For datasheet interpretation, PCB-QA retrieves relevant evidence from chip datasheets to answer device-specific questions~\cite{zhu2026pcbqa}.

Prior systems also show how structured specifications and hardware observations can support executable tasks.
Driver synthesis systems generate implementations from formal device specifications~\cite{wang2003peripheral,ryzhyk2009termite}.
Perry derives MCU peripheral models from existing drivers, while P$^2$IM models peripheral interfaces to execute and test firmware~\cite{lei2024perry,feng2020p2im}.
Avatar combines firmware emulation with real hardware for dynamic analysis~\cite{zaddach2014avatar}.
These efforts motivate a question for datasheet-driven development: how can the device definitions extracted to answer an engineering question also be reused to evaluate behavior, generate a driver, and integrate the device into firmware?

\begin{figure}[t!]
  \centering
  \includegraphics[width=\linewidth]{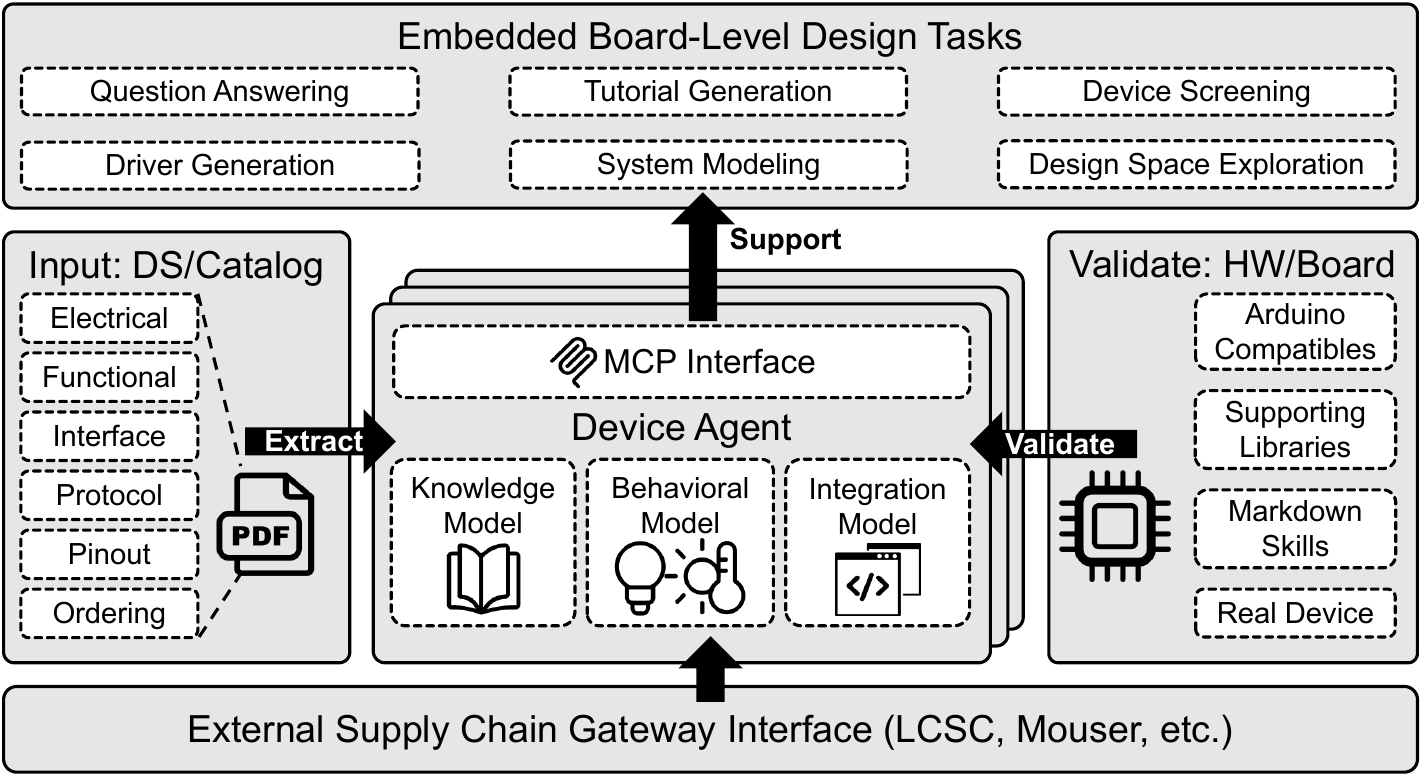}
  \caption{DS2Agent constructs three linked models from a device datasheet and exposes their capabilities through MCP.}
  \label{fig:system}
\end{figure}

We propose \textbf{DS2Agent}, a framework that constructs reusable device agents from manufacturer datasheets.
As shown in Fig.~\ref{fig:system}, each agent comprises three linked models:
\begin{inparaenum}[(1)]
\item a \emph{knowledge model} that supports retrieval and question answering grounded in the datasheet,
\item a \emph{behavior model} that captures documented device behavior, including register semantics and transfer functions, and
\item a \emph{driver model} that represents digital-interface operations and supports portable C driver generation.
\end{inparaenum}
By analogy with a compiler, these models form a shared intermediate representation that downstream tasks can reuse without independently extracting the same definitions from the datasheet.
For example, a device agent can explain how a measurement is encoded, evaluate the conversion for a supplied raw value, and provide the driver operations needed to acquire that value from a device.

DS2Agent exposes these capabilities through the Model Context Protocol (MCP).
Downstream agents can use the interface to retrieve specifications and their sources, evaluate modeled behavior, and obtain drivers for board integration.
This supports early exploration of candidate designs using the documented behavior available for each device.
When a physical device is available, DS2Agent can compare its measurements with model predictions and use the observations to refine instance-specific calibration.
The documented specification remains the reference for these comparisons, and measured values are recorded with the device instance and its execution configuration.
Table~\ref{tbl:compare} summarizes the capabilities of DS2Agent and selected tools.

\head{Contributions.}
\begin{itemize}
  \item \textbf{Device representation.} We design a representation that links knowledge, behavior, and driver models derived from the same datasheet, allowing retrieval, evaluation, and device operations to share source-attributed definitions (\S\ref{sec:design}).
  \item \textbf{Hardware feedback.} We introduce optional real-device exploration that checks modeled behavior against hardware observations and supports instance-specific calibration while preserving documented specifications (\S\ref{sec:impl}).
  \item \textbf{Implementation.} We implement DS2Agent with a multi-agent pipeline that constructs and verifies the three models, generates portable C drivers, and exposes the resulting capabilities through a unified MCP interface (\S\ref{sec:impl}).
  \item \textbf{Evaluation.} We evaluate DS2Agent on ten devices spanning seven functional categories across question answering, driver generation, and system integration (\S\ref{sec:eval}).
  The augmented agent scores 97 out of 100 on the question-answering benchmark, compared with 87 for the baseline, while using 44.8\% fewer tokens.
  It reduces average token usage per device for driver generation by 72.9\% and total token usage for system integration by 29.7\% across three FreeRTOS applications.
\end{itemize}

\begin{figure*}[t!]
  \centering
  \includegraphics[width=\textwidth, trim=0 0 0 0, clip]{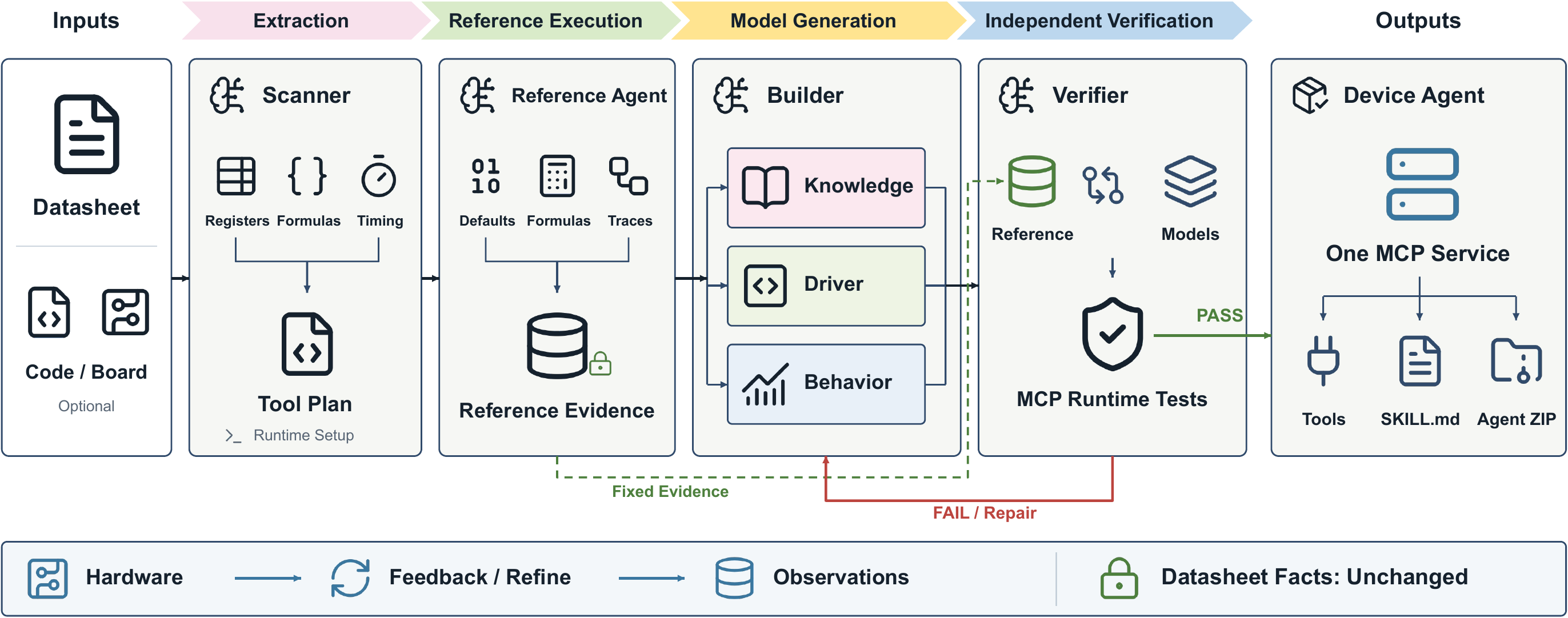}
  \caption{Overview of the ds2agent workflow for generating and validating device agents from datasheets.}
  \label{fig:workflow}
\end{figure*}
\section{Background and Related Work}
\label{sec:related}
\textbf{Artifacts as agent capabilities.}
Paper2Agent~\cite{paper2agent} constructs MCP interfaces around research papers
and their associated codebases, using iterative testing to improve reliability.
We adopt this reusable-artifact perspective but study the boundary between
datasheet statements and a limited executable interpretation. MCP standardizes
the interface; it does not certify a device model's fidelity or the independence
of its tests~\cite{mcp}.

\textbf{Datasheet-to-model generation.}
DS2SC-Agent~\cite{ds2sc} generates SystemC chiplet models using a multi-agent
document-parsing and debugging pipeline. Our implementation instead targets
bounded peripheral transactions, reviewed source evidence, separately supplied
acceptance inputs, and freshness checks before agent access. We do not claim
that automatic modeling itself is new, or that our narrow evaluation establishes
superiority over SystemC generation systems.


\textbf{Virtual platforms and board tools.}
QEMU~\cite{qemu} and Renode~\cite{renode} provide executable virtual platforms;
their existence does not imply that an arbitrary imported device description
has a matching peripheral adapter. KiCad~\cite{kicad} supplies schematic and PCB
tools, but generating files or a placement skeleton is not equivalent to
completing routing, electrical-rule checks or physical acceptance.
Our focus is the auditable transition between evidence, supported behavior and
agent-visible claims, rather than replacing these downstream tools.

\section{Design}
\label{sec:design}
Fig.~\ref{fig:workflow} shows how DS2Agent constructs and verifies a device agent from a datasheet.
DS2Agent constructs three linked models for device knowledge, driver operations, and behavioral semantics, and exposes their verified capabilities through the Model Context Protocol (MCP).
The models supply shared device definitions for knowledge retrieval, behavioral evaluation, and driver execution.
Each operation provides evidence within its own scope.

\subsection{From Datasheets to Three Linked Models}
\label{sec:model_generation}

\paragraph{Canonical device representation}
For each device type $d$, the canonical representation is
\begin{equation}
  \mathcal{M}_d = (\mathcal{K}_d,\mathcal{D}_d,\mathcal{B}_d),
  \label{eq:canonical}
\end{equation}
where $\mathcal{K}_d$, $\mathcal{D}_d$, and $\mathcal{B}_d$ denote the knowledge, driver, and behavior models, respectively.
The three models capture complementary aspects of the same device and reference shared definitions through canonical identifiers.
For example, the driver and behavior models can reference the same register definitions and conversion coefficients.
Generated code, register tables, and explanatory text are derived from these definitions.

\paragraph{Source-grounded construction}
DS2Agent fixes reference evidence derived from documented examples, defaults, and formulas before constructing the candidate models.
An independent verifier runs the candidate's MCP tools and compares their outputs with the reference evidence.
When a check fails, the observed outputs and discrepancies guide the builder's repair.
The verifier runs the repaired candidate again against the same reference evidence.
Device facts retain their source references and applicability conditions.

\paragraph{Feedback and evidence boundaries}
When hardware is available, DS2Agent records the results of board execution as evidence for the physical instance.
These observations can guide repairs to the agent or board integration and support instance-specific calibration.
After a repair, DS2Agent reruns the relevant verification checks.
Source review, model evaluation, native driver execution, target builds, QEMU execution, and physical tests support different verification claims.
Claims about physical behavior and hardware calibration require evidence from physical tests.

\paragraph{Driver model}
The driver model describes documented digital-interface operations, including initialization, configuration, register and command access, acquisition sequences, and required delays.
It references canonical register definitions and conversion parameters and supports generation of portable C driver code.
A transport contract separates device logic from platform-specific bus, GPIO, and timing operations.
Board integration implements this contract through hardware adapters while reusing the portable device driver.

\paragraph{Behavior model}
The behavior model provides a finite, parameterized description of the device, covering pin and supply conditions, register and command semantics, reset values, states and transitions, parameter bounds, and transfer functions.
Each modeled relation retains its documented applicability conditions.
Register checks use masks only where supported by the source, and documented variants may define sets of legal values.
These documented constraints define the acceptance criteria for the device type.

\paragraph{Consistency and instance evidence}
Verification checks that shared definitions and references resolve consistently across the three models.
The representation separates immutable specification facts $\mathcal{I}_d$ from mutable observations and calibrations $\mathcal{O}_{d,i}$ for physical instance $i$.
Under a fixed source revision, refinement must preserve
\begin{equation}
  \mathcal{I}_d^{\mathrm{after}} = \mathcal{I}_d^{\mathrm{before}}.
  \label{eq:immutable}
\end{equation}
Measured values are appended to the instance evidence, while datasheet typical values, formulas, and identity constraints remain in $\mathcal{I}_d$.
An unexpected chip ID is recorded as a deviation and may trigger a review of the source or part identity.
Each accepted refinement retains the device, board, and software configuration under which its supporting observations were obtained.

\subsection{Model-Backed Capabilities and MCP Access}
\label{sec:capabilities}

A single MCP service exposes the verified capabilities and routes requests to the corresponding implementations.

\paragraph{Capability semantics}
Knowledge capabilities retrieve and explain cited facts, behavior capabilities inspect and evaluate modeled relations, and driver capabilities perform documented operations or export the portable driver.
For example, a temperature conversion evaluates the documented relation between a supplied raw code and temperature.
A hardware measurement also acquires the raw code from a bound physical instance before converting it.
Each tool is backed by a source-supported model record or a documented device operation.
Unsupported capabilities are explicitly marked as unavailable.

\paragraph{Board specialization and instance isolation}
Board specialization binds the existing device models and portable driver to a target board through hardware adapters and initialization firmware.
The board binding describes how the transport contract is implemented on that board while preserving the portable driver code.
A shared MCP service can expose several device agents, with dispatch preserving each device's definitions and each physical instance's observations and calibration provenance.
Calibration records apply to the physical instance and configuration under which their supporting evidence was collected.

\paragraph{Feedback and evidence boundaries}
Before hardware is available, the device agent supports source-grounded inspection, behavioral evaluation, and driver preparation.
Once hardware is available, measurements add instance evidence and can support refinement of the mutable layer under Eq.~\eqref{eq:immutable}.
The construction workflow verifies candidate models against fixed reference evidence, while hardware feedback accumulates observations from a physical instance.
Source review, model evaluation, native driver execution, target builds, QEMU execution, and physical tests support different verification claims.
Claims about physical behavior and hardware calibration require evidence from physical tests.
\section{Implementation}
\label{sec:impl}
\subsection{Agent Packaging and Model Storage}
\label{sec:implementation_storage}

DS2Agent delivers each verified agent as a ZIP archive containing the public MCP entry point \texttt{agent\_mcp.py}, \texttt{SKILL.md}, package metadata, and dependency specifications.
Scope-wide capability descriptions are stored under \texttt{capabilities/}.
Each device directory under \texttt{devices/} contains the three canonical models, the device module, the portable driver, and native test support.
For board integration, a corresponding directory under \texttt{boards/} stores the board binding, firmware, and build artifacts.

Each model records its \texttt{device\_id}, \texttt{model\_revision}, supported and unsupported scope, and source references.
The knowledge model stores cited facts and their applicability conditions.
The driver model records initialization, configuration, register access, delays, data interpretation, restrictions, and transaction structure.
Ordered sequences of operations encoded in JSON specify the initialization, configuration, acquisition, and interpretation flows.
The behavior model records register and command semantics, state transitions, minimum, typical, and maximum parameter values, and transfer functions.
The models reference shared register definitions, defaults, and coefficients through canonical identifiers.

\subsection{Multi-Agent Construction Workflow}
\label{sec:implementation_workflow}

The construction workflow in Fig.~\ref{fig:workflow} assigns extraction, reference execution, model generation, and verification to separate roles.
During extraction, the datasheet scanner and environment manager run concurrently to prepare the tool plan and runtime setup.
Reference executors evaluate selected examples and formulas to produce expected values and reference traces where applicable.
The workflow fixes this evidence before model generation begins.

Each implementation module has one assigned implementer.
An independent verifier checks the candidate models and exercises their MCP tools against the fixed reference evidence.
A failed check returns the affected module to its implementer for repair, after which the verifier checks the revision against the same evidence.
DS2Agent packages the agent after verification passes.
Phase barriers enforce prerequisites, and an agent-run record captures phase execution, module ownership, and the separation between implementers and verifiers.

\subsection{Portable Driver and Board Integration}
\label{sec:implementation_driver}

The generated driver interacts with hardware through \texttt{write}, \texttt{read}, \texttt{gpio}, and \texttt{delay} callbacks.
The driver is independent of MCU and vendor HAL headers and leaves peripheral access and the application entry point to the board layer.
Board adapters implement the callbacks and handle bus bindings, address conventions, chip-select control, timeouts, and locking as required.
Board initialization configures shared peripherals and creates device instances with explicit bus, address, and control-pin bindings.

Adapters preserve the device's documented transaction semantics.
When the target HAL expects a shifted I$^2$C address, the adapter shifts the canonical 7-bit address exactly once.
A combined register-read operation such as \texttt{Mem\_Read} is used only when its repeated-START sequence matches the device protocol.
For a documented status-byte protocol, the driver first reads the status and, if \texttt{RDY=1}, reads the frame in a separate transfer with a length derived from \texttt{LEN}.
For each interrupt used, integration checks that the peripheral interrupt is enabled, the corresponding NVIC entry is enabled, and the ISR invokes the matching handler.

The board binding in \texttt{boards/<board\_id>/board.json} records the board model, hardware revision, and MCU identity with source references, together with the pin, clock, bus, and control plans and their hashes.
It identifies artifacts by role, including initialization code, HAL code, application and control sources, the build recipe, startup code, and the linker script.
The binding records target-build and hardware-initialization evidence separately.
The \texttt{binding\_sha256} field hashes the canonical binding with output hashes and reports excluded to avoid a circular dependency.
Board specialization reuses the portable driver byte for byte, and its content hash tracks that identity across board migration, merging, and relocation.

\subsection{MCP Dispatch and Capability Projections}
\label{sec:implementation_mcp}

A single MCP server exposes the supported capabilities of the validated device modules.
Tool names include a device or board identifier, allowing the server to route each request to the corresponding implementation.
The implementation resolves the relevant canonical records or bound execution context.
Capability descriptions map tasks to available tools and identify unsupported operations.
Board tools use separate board bindings to support driver integration, initialization, and state inspection.

\subsection{Feedback Provenance and Verification}
\label{sec:implementation_verification}

Hardware runs record measurements and gaps in model coverage as instance evidence.
Calibration and refinement update only mutable records, and snapshot comparisons enforce Eq.~\eqref{eq:immutable}.
Each accepted refinement is stored with its physical instance record and records the configuration under which the supporting evidence was obtained:
\begin{equation}
  (d, b, h_{\mathrm{driver}}, h_{\mathrm{HAL}}, h_{\mathrm{plan}}),
  \label{eq:refinement_identity}
\end{equation}
where $d$ and $b$ are the device and board identifiers, and the remaining entries are SHA-256 hashes of the driver, HAL, and execution plan, respectively.

Register readbacks use exact equality by default.
Where documented, integer checks may apply a bit mask or test membership in a set of legal values.
Floating-point comparisons allow deviations only within a finite tolerance supported by the source.

Three verification programs check workflow integrity, MCP consistency, and board evidence.
Native tests compile and run the unmodified exported driver against a fake transport to check conformance to the modeled transaction rules.
Target builds, QEMU runs, and physical tests assess buildability, emulated execution, and behavior on actual hardware, respectively.
\section{Evaluation}
\label{sec:eval}
We evaluate our design across three core tasks: question answering, driver generation, and system integration.
All tasks are executed using Oh My Pi~\cite{ohmypi2026} (v18.18.2), an open-source coding agent/harness that supports task delegation to subagents, with DeepSeek-V4.1-Flash~\cite{deepseek2026} as the underlying language model.
We compare two agent configurations: a DS2Agent-augmented agent equipped with DS2Agent-generated artifacts and a baseline agent without these artifacts.
We use the NUCLEO-F103RB~\cite{st_nucleo_f103rb} as the target board and evaluate the following ten devices:
\begin{itemize}
    \item SHT31: temperature and humidity sensor;
    \item ADXL345: three-axis accelerometer;
    \item QMC5883L: three-axis magnetometer;
    \item HX711: 24-bit ADC for load cells;
    \item INA219: current, voltage, and power monitor;
    \item MCP4725: 12-bit digital-to-analog converter;
    \item PCA9548A: eight-channel I$^2$C bus switch;
    \item AT24C256: EEPROM;
    \item DS3231: temperature-compensated real-time clock;
    \item PCF8574T: eight-bit I$^2$C I/O expander.
\end{itemize}
The ten devices span seven functional categories: sensors, data converters, electrical monitors, bus switches, memory, real-time clocks, and I/O expanders.
For the system integration task, we use FreeRTOS V11.3.3~\cite{freertos} as the real-time operating system.

\subsection{Question Answering}

\begin{table}[t]
\centering
\caption{Question-answering scores by category.}
\label{tbl:qa_scores_by_category}
\scriptsize
\setlength{\tabcolsep}{4pt}
\renewcommand{\tabularxcolumn}[1]{m{#1}}
\begin{tabularx}{\columnwidth}{|>{\raggedright\arraybackslash}X|c|c|c|}
\hline
\textbf{Question Category} &
\textbf{Questions} &
\makecell[cc]{\textbf{Augmented}\\\textbf{Agent Score}} &
\makecell[cc]{\textbf{Baseline}\\\textbf{Agent Score}} \\
\hline
Communication & 20 & 20 & 19 \\
\hline
Power Consumption and Supply & 20 & 20 & 18 \\
\hline
Functionality & 20 & 20 & 20 \\
\hline
Timing & 20 & 20 & 16 \\
\hline
Electrical Characteristics and Interference & 10 & 8 & 7 \\
\hline
Reliability and Environmental Conditions & 10 & 9 & 7 \\
\hline
\textbf{Total} & \textbf{100} & \textbf{97} & \textbf{87} \\
\hline
\end{tabularx}
\end{table}

We evaluate both agents on the same 100 questions, with 10 questions per device across the six categories listed in Table~\ref{tbl:qa_scores_by_category}.
For each agent, we record total token usage, total response time, and answer scores.

The augmented agent uses 1,021,312 tokens and takes 137~s, compared with 1,849,354 tokens and 197~s for the baseline.
This corresponds to a 44.8\% reduction in token usage and a 30.5\% reduction in total response time.
As shown in Table~\ref{tbl:qa_scores_by_category}, the augmented agent scores 97 out of 100, compared with 87 for the baseline.(Table~\ref{tbl:qa_scores_by_category}).
The largest gain occurs in the timing category, where the augmented agent scores 20 out of 20, compared with 16 for the baseline.

Both agents use the same datasheets converted to Markdown.
DS2Agent extracts and organizes information from those datasheets into structured models that the augmented agent can query.
On this benchmark, access to these models is associated with both lower answering cost and higher scores.

\subsection{Driver Generation}

For each of ten devices, both agents were tasked with implementing the
same specified functionality on the NUCLEO-F103RB and producing a
compiled binary. The baseline agent worked from the device datasheet,
converted from PDF to Markdown using MinerU~\cite{wang2024mineru}, and the provided SDK.
Its task instructions prohibited use of the \texttt{ds2agent} skill.
The augmented agent accessed device-specific artifacts, including a pre-generated portable C driver through the corresponding prebuilt device-agent MCP server.

Datasheet conversion and device-agent preparation took place before
the timed runs. For each run, we recorded token usage and measured
wall-clock time from task submission to the first successful target
build, including compilation time.

Table~\ref{tbl:driver_generation} reports the per-device results.
Across the ten devices, the augmented agent reduced mean token usage
by 72.9\% and mean downstream task completion time from 737.2\,s to
24.86\,s, a reduction of 96.6\%.

\begin{table}[t!]
\scriptsize
\centering
\caption{\small Driver generation comparison across ten devices.}
\label{tbl:driver_generation}
\begin{tabular}{|l|r|c|r|c|}
  \hline
  \multicolumn{1}{|c|}{\multirow{2}{*}{\textbf{Device}}} &
  \multicolumn{2}{c|}{\textbf{Augmented agent}} &
  \multicolumn{2}{c|}{\textbf{Baseline agent}} \\
  \cline{2-5}
  & Token usage & Time (s) & Token usage & Time (s) \\
  \hline
  SHT31    & 2,258,912 & 20   & 6,633,968  & 800 \\
  \hline
  ADXL345  & 2,714,985 & 31   & 9,194,883  & 648 \\
  \hline
  QMC5883L & 2,801,481 & 30   & 9,917,133  & 818 \\
  \hline
  HX711    & 2,424,998 & 25   & 12,362,224 & 968 \\
  \hline
  INA219   & 2,975,487 & 22   & 15,647,209 & 489 \\
  \hline
  MCP4725  & 3,045,502 & 24   & 7,736,161  & 470 \\
  \hline
  PCA9548A & 2,646,912 & 26   & 6,003,804  & 583 \\
  \hline
  AT24C256 & 3,118,262 & 25   & 12,625,547 & 679 \\
  \hline
  DS3231   & 3,182,346 & 23   & 19,402,812 & 1,307 \\
  \hline
  PCF8574T & 3,445,152 & 23   & 6,108,769  & 610 \\
  \hline
\end{tabular}
\end{table}

\subsection{System Integration}

We evaluate the DS2Agent-augmented agent and the baseline agent on three
FreeRTOS system integration tasks. Both agents receive the same application
requirements. For each task, we measure token usage and completion time.
Each application includes a default task that logs boot messages and
handles commands over UART.

\begin{figure}[t!]
  \centering
  \includegraphics[width=\linewidth, trim=12 15 5 18, clip]{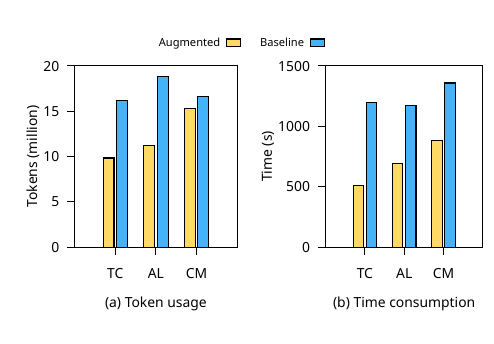}
  \caption{Token usage and completion time for the three FreeRTOS tasks.}
  \label{fig:ds2a_is}
\end{figure}

\paragraph{Temperature control (TC)}
A monitoring task samples the SHT31 once per second. When the measured
temperature exceeds a configured threshold, it notifies a fan-control
task, which turns on the fan.

\paragraph{Automatic leveling (AL)}
A monitoring task samples the ADXL345 every 20~ms to estimate platform
tilt. When the absolute tilt exceeds $5^\circ$, it notifies a
motor-control task to drive the platform toward level. Corrective motion
stops when the absolute tilt falls below $2^\circ$. The two thresholds
provide hysteresis and limit repeated motor switching near the activation
threshold. The application reports the estimated tilt and control status
over UART.

\paragraph{Current monitoring and load disconnection (CM)}
A monitoring task samples the INA219 every 100~ms. When the measured
current exceeds a configured threshold, it notifies a relay-control task,
which opens the relay to disconnect the load. The application reports the
measured current and relay status over UART.

\paragraph{Results}
The implementations produced by both agents passed the functional checks
for all three tasks. We verified fan activation, tilt correction and
stopping, and load disconnection under the specified conditions.
As shown in Fig.~\ref{fig:ds2a_is}, the DS2Agent-augmented agent used fewer
tokens and completed each task faster than the baseline. Across the three
tasks, it reduced total token usage by 29.7\% and total completion time
by 44.1\%. Per-task time reductions ranged from 35.0\% to 57.5\%.
Temperature control showed the largest time reduction, from 1,199~s
to 510~s, while automatic leveling showed the largest token reduction
at 40.6\%.
\section{Conclusion}
\label{sec:conc}
We presented DS2Agent, a framework that transforms device datasheets into reusable device agents for question answering, driver generation, and system integration.
DS2Agent organizes device knowledge, driver operations, and behavioral semantics into three linked models and exposes them through MCP tools, allowing downstream agents to reuse the same device definitions across tasks.

On the 100-question benchmark, the DS2Agent-augmented agent scored 97 out of 100, compared with 87 for the baseline, while using 44.8\% fewer tokens and taking 30.5\% less time.
For driver generation on the NUCLEO-F103RB, it reduced average token usage by 72.9\% and average completion time by 96.6\% across ten devices.
Across the three FreeRTOS integration tasks, it reduced total token usage by 29.7\% and total completion time by 44.1\%.

These results suggest that compiling datasheets into reusable device models can reduce the effort required for embedded system design.

\clearpage
\bibliographystyle{IEEEtran}
\bibliography{refs}

@misc{paper2agent,
  author = {Jiacheng Miao and Joe R. Davis and Yaohui Zhang and Jonathan K. Pritchard and James Zou},
  title = {{Paper2Agent}: Reimagining Research Papers As Interactive and Reliable {AI} Agents},
  year = {2025},
  eprint = {2509.06917},
  archivePrefix = {arXiv},
  doi = {10.48550/arXiv.2509.06917},
  url = {https://arxiv.org/abs/2509.06917}
}

@misc{mcp,
  author = {{Model Context Protocol Contributors}},
  title = {{Model Context Protocol} Specification, Revision 2025-03-26},
  year = {2025},
  howpublished = {Online specification},
  url = {https://modelcontextprotocol.io/specification/2025-03-26}
}

@misc{ds2sc,
  author = {Yiwei Wu and Yifan Wu and Yunhao Xiong and Dengwei Zhao and Jiaxuan Shen and Jianfei Jiang and Guanghui He and Shikui Tu and Yanan Sun},
  title = {{DS2SC-Agent}: A Multi-Agent Automated Pipeline for Rapid Chiplet Model Generation},
  year = {2026},
  eprint = {2603.21190},
  archivePrefix = {arXiv},
  doi = {10.48550/arXiv.2603.21190},
  url = {https://arxiv.org/abs/2603.21190}
}

@inproceedings{qemu,
  author = {Fabrice Bellard},
  title = {{QEMU}, a Fast and Portable Dynamic Translator},
  booktitle = {Proceedings of the USENIX Annual Technical Conference, FREENIX Track},
  pages = {41--46},
  year = {2005},
  url = {https://www.usenix.org/legacy/events/usenix05/tech/freenix/bellard.html}
}

@misc{renode,
  author = {{Antmicro}},
  title = {{Renode} Documentation},
  howpublished = {Online documentation},
  url = {https://renode.readthedocs.io/}
}

@misc{kicad,
  author = {{KiCad Project}},
  title = {{KiCad} Documentation},
  howpublished = {Online documentation},
  url = {https://docs.kicad.org/}
}

@misc{arduinoMCP,
  author       = {{Hardware MCP}},
  title        = {{Arduino MCP Server}},
  howpublished = {\url{https://github.com/hardware-mcp/arduino-mcp-server}},
  year         = {2026},
  note         = {Open-source GitHub repository}
}

@misc{embeddedMCP,
  author       = {Atilla},
  title        = {{Embedded MCP}},
  howpublished = {\url{https://github.com/atillab1/embedded-mcp}},
  year         = {2026},
  note         = {Open-source GitHub repository}
}

@misc{labgrid,
  author       = {{Labgrid Project}},
  title        = {{Labgrid: Embedded Systems Control Library}},
  howpublished = {\url{https://github.com/labgrid-project/labgrid}},
  year         = {2026}
}

@misc{eclipseDitto,
  author       = {{Eclipse Foundation}},
  title        = {{Eclipse Ditto: Digital Twin Framework}},
  howpublished = {\url{https://github.com/eclipse-ditto/ditto}},
  year         = {2026}
}

@article{openCalib,
  author  = {Yan, Guohang and others},
  title   = {{OpenCalib: A Multi-sensor Calibration Toolbox for Autonomous Driving}},
  journal = {arXiv preprint arXiv:2205.14087},
  year    = {2022},
  url     = {https://github.com/Cartesian123/OpenCalib}
}

@misc{ohmypi2026,
  author       = {B{\"o}l{\"u}k, Can},
  title        = {{Oh My Pi} (Version 18.18.2)},
  year         = {2026},
  howpublished = {\url{https://github.com/can1357/oh-my-pi}},
  note         = {Accessed November 15, 2026}
}

@misc{deepseek2026,
  author       = {{DeepSeek}},
  title        = {{DeepSeek-V4.1-Flash}: Smarter, Faster, More Efficient},
  year         = {2026},
  month        = sep,
  howpublished = {\url{https://api-docs.deepseek.com/news/news260910/}},
  note         = {Accessed November 15, 2026}
}

@manual{st_nucleo_f103rb,
  author = {{STMicroelectronics}},
  title  = {{STM32 Nucleo-64 Boards (MB1136)}},
  year   = {2025},
  month  = sep,
  note   = {User manual UM1724, Rev. 17},
  url    = {https://www.st.com/resource/en/user_manual/um1724-stm32-nucleo64-boards-mb1136-stmicroelectronics.pdf}
}

@misc{freertos,
  author = {{FreeRTOS}},
  title  = {{FreeRTOS Kernel V11.3.3}},
  year   = {2026},
  month  = oct,
  url    = {https://github.com/FreeRTOS/FreeRTOS-Kernel/releases/tag/V11.3.3},
  note   = {Accessed: Nov. 15, 2026}
}

@misc{liu2023chipnemo,
  title         = {{ChipNeMo}: Domain-Adapted {LLMs} for Chip Design},
  author        = {Liu, Mingjie and Ene, Teodor-Dumitru and Kirby, Robert and others},
  year          = {2023},
  eprint        = {2311.00176},
  archivePrefix = {arXiv},
  primaryClass  = {cs.CL},
  doi           = {10.48550/arXiv.2311.00176},
  url           = {https://arxiv.org/abs/2311.00176}
}

@inproceedings{lin2020polymorphic,
  author = {Lin, Richard and Ramesh, Rohit and Chi, Connie and Jain, Nikhil and Nuqui, Ryan and Dutta, Prabal and Hartmann, Bj{\"o}rn},
  title = {Polymorphic Blocks: Unifying High-Level Specification and Low-Level Control for Circuit Board Design},
  booktitle = {Proceedings of the 33rd Annual ACM Symposium on User Interface Software and Technology},
  year = {2020},
  pages = {529--540},
  doi = {10.1145/3379337.3415860}
}

@inproceedings{lin2024boarddse,
  author = {Lin, Richard and Ramesh, Rohit and Pandhare, Parth and Tay, Kai Jun and Dutta, Prabal and Hartmann, Bj{\"o}rn and Mehta, Ankur},
  title = {Design Space Exploration for Board-level Circuits: Exploring Alternatives in Component-based Design},
  booktitle = {Proceedings of the CHI Conference on Human Factors in Computing Systems},
  year = {2024},
  doi = {10.1145/3613904.3642009}
}

@article{zhu2026pcbqa,
  author = {Zhu, Binwu and Yu, Bei},
  title = {{PCB-QA}: A Task-Specific Question Answering Framework for Text-Centric Chip Datasheet Queries in {PCB} Design},
  journal = {ACM Transactions on Design Automation of Electronic Systems},
  year = {2026},
  doi = {10.1145/3811922}
}

@inproceedings{li2026pcbbench,
  author = {Li, Jindong and Chen, Lianrong and Yang, Bin and Zhu, Jiadong and Wang, Ying and Ma, Yuzhe and Yang, Menglin},
  title = {{PCB-Bench}: Benchmarking {LLMs} for Printed Circuit Board Placement and Routing},
  booktitle = {International Conference on Learning Representations},
  year = {2026},
  url = {https://proceedings.iclr.cc/paper_files/paper/2026/hash/051f3997af1dd65da8e14397b6a72f8e-Abstract-Conference.html}
}

@inproceedings{chen2025pcbagent,
  author = {Chen, Lin and Chen, Ran and Hu, Shoubo and Yao, Xufeng and Tang, Zhentao and Kai, Shixiong and Xu, Siyuan and Yuan, Mingxuan and Hao, Jianye and Yu, Bei and Xu, Jiang},
  title = {{PCBAgent}: An Agent-based Framework for High-Density Printed Circuit Board Placement},
  booktitle = {Proceedings of the 30th Asia and South Pacific Design Automation Conference},
  year = {2025},
  pages = {781--787},
  doi = {10.1145/3658617.3697736}
}

@article{liu2025rtlcoder,
  author = {Liu, Shang and Fang, Wenji and Lu, Yao and Wang, Jing and Zhang, Qijun and Zhang, Hongce and Xie, Zhiyao},
  title = {{RTLCoder}: Fully Open-Source and Efficient {LLM}-Assisted {RTL} Code Generation Technique},
  journal = {IEEE Transactions on Computer-Aided Design of Integrated Circuits and Systems},
  year = {2025},
  volume = {44},
  number = {4},
  pages = {1448--1461},
  doi = {10.1109/TCAD.2024.3483089}
}

@inproceedings{wang2003peripheral,
  author = {Wang, Shaojie and Malik, Sharad and Bergamaschi, Reinaldo A.},
  title = {Modeling and Integration of Peripheral Devices in Embedded Systems},
  booktitle = {Proceedings of the Design, Automation and Test in Europe Conference and Exhibition},
  year = {2003},
  doi = {10.1109/DATE.2003.1253599}
}

@inproceedings{ryzhyk2009termite,
  author = {Ryzhyk, Leonid and Chubb, Peter and Kuz, Ihor and Le Sueur, Etienne and Heiser, Gernot},
  title = {Automatic Device Driver Synthesis with {Termite}},
  booktitle = {Proceedings of the ACM SIGOPS 22nd Symposium on Operating Systems Principles},
  year = {2009},
  pages = {73--86},
  doi = {10.1145/1629575.1629583}
}

@inproceedings{lei2024perry,
  author = {Lei, Chongqing and Ling, Zhen and Zhang, Yue and Yang, Yan and Luo, Junzhou and Fu, Xinwen},
  title = {A Friend's Eye is A Good Mirror: Synthesizing {MCU} Peripheral Models from Peripheral Drivers},
  booktitle = {33rd USENIX Security Symposium},
  year = {2024},
  pages = {7085--7102},
  url = {https://www.usenix.org/conference/usenixsecurity24/presentation/lei}
}

@inproceedings{feng2020p2im,
  author = {Feng, Bo and Mera, Alejandro and Lu, Long},
  title = {{P2IM}: Scalable and Hardware-independent Firmware Testing via Automatic Peripheral Interface Modeling},
  booktitle = {29th USENIX Security Symposium},
  year = {2020},
  pages = {1237--1254},
  url = {https://www.usenix.org/conference/usenixsecurity20/presentation/feng}
}

@inproceedings{zaddach2014avatar,
  author = {Zaddach, Jonas and Bruno, Luca and Francillon, Aur{\'e}lien and Balzarotti, Davide},
  title = {{Avatar}: A Framework to Support Dynamic Security Analysis of Embedded Systems' Firmwares},
  booktitle = {Network and Distributed System Security Symposium},
  year = {2014},
  doi = {10.14722/ndss.2014.23229}
}

@article{wang2024mineru,
  author  = {Wang, Bin and Xu, Chao and Zhao, Xiaomeng and Ouyang, Linke
             and Wu, Fan and Zhao, Zhiyuan and Xu, Rui and Liu, Kaiwen
             and Qu, Yuan and Shang, Fukai and Zhang, Bo and Wei, Liqun
             and Sui, Zhihao and Li, Wei and Shi, Botian and Qiao, Yu
             and Lin, Dahua and He, Conghui},
  title   = {{MinerU}: An Open-Source Solution for Precise Document Content Extraction},
  journal = {arXiv preprint arXiv:2409.18839},
  year    = {2024}
}

\end{document}